\documentclass{vgtc}                          

\ifpdf
  \pdfoutput=1\relax                   
  \usepackage{graphicx}                
  \DeclareGraphicsExtensions{.pdf,.png,.jpg,.jpeg} 
\else
  \usepackage{graphicx}                
  \DeclareGraphicsExtensions{.eps}     
\fi%

\graphicspath{{figures/}{pictures/}{images/}{./}} 

\usepackage{microtype}                 
\PassOptionsToPackage{warn}{textcomp}  
\usepackage{textcomp}                  
\usepackage{mathptmx}                  
\usepackage{times}                     
\usepackage{cite}                      
\usepackage{tabu}                      
\usepackage{wrapfig}
\usepackage{booktabs}   
\usepackage{booktabs}
\usepackage{soul}
\usepackage{color}
\usepackage{amssymb}
\usepackage{amsmath}
\DeclareMathOperator*{\argmax}{argmax}
\usepackage{algorithm,algorithmic}
\usepackage{subfigure}

\onlineid{5799}

\vgtccategory{Research}

\vgtcinsertpkg

\newcommand{\name}{PolicyExplainer}
\newcommand{\person}{Gary}

\title{Why? Why not? When? Visual Explanations of Agent Behaviour in Reinforcement Learning}

\author{Aditi Mishra\thanks{e-mail: amishr45@asu.edu} %
\and Utkarsh Soni\thanks{e-mail:usoni1@asu.edu} %
\and Jinbin Huang\thanks{e-mail:jhuan196@asu.edu}
\and Chris Bryan\thanks{e-mail:cbryan16@asu.edu}}

\abstract{Reinforcement learning (RL) is used in many domains, including autonomous driving, robotics, stock trading, and video games. Unfortunately, the black box nature of RL agents, combined with legal and ethical considerations, makes it increasingly important that humans (including those are who not experts in RL) understand the reasoning behind the actions taken by an RL agent, particularly in safety-critical domains. To help address this challenge, we introduce \textit{\name{}}, a visual analytics interface which lets the user directly query an autonomous agent. \name{} visualizes the states, policy, and expected future rewards for an agent, and supports asking and answering questions such as: ``\textit{Why take this action? Why not take this other action? When is this action taken?}'' \name{} is designed based upon a domain analysis with RL researchers, and is evaluated via qualitative and quantitative assessments on a trio of domains: taxi navigation, a stack bot domain, and drug recommendation for HIV patients.%
We find that \name{}'s visual approach promotes trust and understanding of agent decisions better than a state-of-the-art text-based explanation approach. Interviews with domain practitioners provide further validation for \name{} as applied to safety-critical domains. Our results help demonstrate how visualization-based approaches can be leveraged to decode the behavior of autonomous RL agents, particularly for RL non-experts.
} 

\CCScatlist{
  \CCScatTwelve{Human-centered computing}{Visu\-al\-iza\-tion}{Visu\-al\-iza\-tion techniques}{Treemaps};
  \CCScatTwelve{Human-centered computing}{Visu\-al\-iza\-tion}{Visualization design and evaluation methods}{}
}

\begin{document}


\maketitle
\section{Introduction}
Reinforcement learning (RL) has become a widely-used technique for training autonomous agents. The ability of RL agents to learn sophisticated decision-making in uncertain and complex environments has led to widespread application and success, including in video gaming, autonomous driving, robotics, healthcare, finance, smart grids, and education~\cite{li2017deep}.



Unfortunately, as artificial intelligence (AI) and machine learning (ML) are increasingly deployed in safety-critical domains, there are emerging legal and ethical concerns due to the black box nature of models~\cite{abel2016reinforcement,whittlestone2021societal}. For example, when a healthcare model recommends a drug treatment plan for a patient (e.g.,~\cite{komorowski2018artificial}), can such recommendations be trusted? In human-robot collaborative environments, a human's reasoning for a decision might differ from a model's. Even when the same decision is reached, model recommendations are inherently untrustworthy without sufficient justification or explanation. This problem is well known in the AI/ML community~\cite{lipton2018mythos}, particularly since there exists no common language for a model to communicate decisions to the human and vice versa~\cite{hayes2017improving,fox2017explainable}.


Particularly for RL models that are being applied for decision-making in safety-critical domains, it is necessary that such agents are answerable to people who potentially have little or no AI/ML expertise. This motivates the current work, where we introduce \textit{\name{}}, a novel visual analytics system for policy explanation to RL non-experts. 

To our knowledge, \name{} represents \textit{the first visual analytics system that supports the direct visual querying of and explanation from an RL agent to non-expert users}. \name{} supports users directly querying an RL agent via three of the most common RL policy questions~\cite{lim2009and, hayes2017improving}: ``\textit{Why take this action? Why not take this other action? When is this action taken?}'' \name{} is motivated based on a pre-study with AI/ML researchers, and is intended to be a first step for general-purpose interfaces for RL agent querying and explanation. In contrast to existing visual analytics interfaces for RL explanation~\cite{mcgregor2017interactive, wang2018dqnviz, jaunet2020drlviz, he2020dynamicsexplorer}, \name{} is model-independent and supports both model-based and model-free algorithms. To evaluate \name{}, we conduct an empirical study with RL non-experts on three domains: the popular Taxi domain~\cite{dietterich2000hierarchical}, an HIV drug recommendation domain~\cite{adams2004dynamic}, and a robot stacking boxes in an industrial environment (StackBot) domain. The results indicate that \name{}'s visual explanation approach for agent question-and-answering is effective, particularly compared to text-based policy explanations created via a state-of-the-art natural language generation technique~\cite{hayes2017improving}.

Succinctly, the contributions of this paper include the following. (1) We analyze design requirements for RL policy visualization and explanation for non-experts, based on a pre-study with with RL researchers and reviewing recent AI/ML literature. (2) We define an explanation generation methodology for visual policy explanation in the form of \textit{Why?}, \textit{Why not?}, and \textit{When?} questions. (3) We develop \name{}, a visual analytics interface that lets a user interactively query an RL agent and provides visual explanations of a policy. (4) Based on our experience in creating and extensively evaluating \name{}, we discuss how visualization-based explanations can increase user trust while lessening the cognitive effort required to understand the decision-making process of an RL agent, particularly for RL non-experts in safety-critical domains.

\vspace{-2mm}
\section{Background on Reinforcement Learning}
\label{sec:background}



Reinforcement learning is a technique to train an autonomous \textbf{agent}, in which the agent interacts with the environment and learns to achieve some desired \textbf{goal} through trial-and-error. In contrast to supervised learning, the agent does not require a training set of labeled examples for the desired behavior; likewise, RL differs from unsupervised learning by not simply learning patterns from unlabeled data. Rather, the agent learns the desired behavior using its own experience interacting with the environment. An agent's overall goal can be defined in terms of a special signal called a \textbf{reward} that the agent gets for taking some action in the environment. Informally, the agent is tasked with learning a behavior that would maximize the total amount of reward it receives in the long term. Concretely, the problem of reinforcement learning can be formalized in terms of a \textbf{Markov decision process} (MDP).

\subsection{Markov Decision Process}
\label{sec:mdp}

MDP formulates a sequential decision-making task where, at each time step, the agent observes the current \textbf{state} of the environment and decides to take an \textbf{action}. This transitions the environment to its next state and the agent receives a reward. The agent keeps acting in the environment until it reaches a terminal state at some time step $T$. The reward the agent obtains for each step is discounted with a discount factor $\gamma$. The total reward obtained by the agent for the sequence of transitions is subsequently defined as the \textbf{return}, $G = \sum_{t=1}^{T}\gamma^{t-1}R_{t}$ where $R_{t}$ is the reward obtained for the $t^{\text{th}}$ transition.

Formally, a MDP is defined as a tuple $\langle \mathcal{S}, \mathcal{A}, \mathcal{T}, \mathcal{R}, \gamma \rangle$ where $\mathcal{S}$ is the set of all possible states of the environment, $\mathcal{A}$ is the set of all possible actions the agent can take at any state, $\mathcal{T} \colon \mathcal{S} \times \mathcal{A} \times \mathcal{S} \to [0, 1]$ is the transition function where $\mathcal{T}(s, a, s')$ gives the probability that the environment will transition to a state $s'$ when the agent takes an action $a$ in state $s$, $\mathcal{R} \colon \mathcal{S} \times \mathcal{A} \times \mathcal{S} \to \mathbb{R}$ is the reward function that gives the reward the agent obtains when it takes an action $a$ in state $s$ causing a transition to state $s'$, and $\gamma$ is the discounting factor applied to the obtained rewards. 

Given an MDP, a \textbf{policy} $\pi$ maps each state to some action $a \in \mathcal{A}$. The \textbf{value} for any state is then the expected return the agent gets when it follows the policy $\pi$ starting from the state. The goal of the agent is to learn a policy, referred to as the \textbf{optimal policy} $\pi^{*}$, that maximizes the value for each state. Lastly, the $Q$-value function, $Q_{\pi}(s, a)$ gives the value obtained obtained if the agent takes the action $a$ in state $s$ and then follow the policy $\pi$. If all the components of the MDP are known to the agent, then it can learn $\pi^{*}$ as $\pi^{*}(s) = \argmax_a Q^{*}(s, a)$, where the $Q^{*}(s, a)$ is obtained by solving the following Bellman optimality equation using dynamic programming algorithms like value iteration or policy iteration\cite{sutton2018reinforcement} : $Q^{*}(s,a) = \sum_{s'} \mathcal{T}(s,a,s')[\mathcal{R}(s,a,s') + \gamma .max_{a}Q^{*}(s',a)]$


\subsection{Interpreting State Features}
\label{sec:interpresting_features}

In this work, we assume each state $s$ of the environment can be expressed as a feature vector $\langle f_1, f_2, \dots, f_n\rangle$ where each feature $f_i$ can be understood by the domain expert that will be using \name{} (this is in line with the policy explanation technique presented in~\cite{hayes2017improving}, where states can be described in terms of binary features via classifiers). For example, for the HIV drug recommendation domain, states are defined via features like \textsf{the number of infected lymphocytes}, \textsf{immune response}, etc. \name{} provides explanations in terms of these features. For simplicity, we pick domains where the states were already defined in terms of features that a human can understand. Hence, the state features that the agent views during interaction with the environment would be the same as the one used for explanations. However, this is not a requirement for our system to work. See Section~\ref{sec:discussion} for discussion about relaxing this requirement in future work.

\subsection{Model-Based and Model-Free Learning}


RL algorithms can be classified as either \textbf{model-based} and \textbf{model-free}~\cite{sutton2018reinforcement}. In model-based RL, the agent learns the model of the environment (specifically, it learns the model components of the MDP incorporating the environment dynamics) and uses that to derive the optimal policy. Conversely, model-free RL algorithms do not require a learned model to obtain the optimal policy.

In this work, we use algorithms from both the classes to train our agent for different domains. For the model-based approach, we assume access to the state set $\mathcal{S}$, and employ a simple sampling strategy to learn the model parameters $\mathcal{T}$ and $\mathcal{R}$. With this strategy, for each state $s \in S$, each action $a \in A$ is executed $k$ times. The value of $\mathcal{T}(s,a,s')$ is then set to $k'/k$ where $k'$ is the number of times the environment transitions to state $s'$ when the agent took the action $a$ in state $s$. The value of $\mathcal{R}(s,a,s')$ is set as the average reward obtained for the transitions $\langle s, a, s'\rangle$. Once the model is learned, we compute the optimal policy using policy iteration.

For the model-free approaches, we use function approximation based methods that learn the Q-function directly from agent's experience. The learning involves the agent interacting with the environment over several episodes improving its Q-function estimate. In each episode, the agent starts at some random initial state and follows an $\epsilon$-greedy policy in which the agent chooses an action that maximizes its $Q$ value with a probability of $1-\epsilon$ or chooses to do a random action (uniformly sampled) with probability $\epsilon$.  
The agent collects experience using the $\epsilon$-greedy policy and then uses it to approximate the $Q$-function. For our domains, we used linear function and neural network based approximations. The former technique represents the $Q$-function as a weighted linear function of features $f_i$ defined over the state:
$
    Q_{\theta}(s, a) = \theta_{1}.f_{1}(s) + \theta_{2}.f_{2}(s) + \dots + \theta_{n}.f_{n}(s)
$
The weights are updated after each individual interaction with the environment, where the agent takes an action $a$ in state $s$ resulting in it transitioning to state $s'$ getting a reward $r$, using the following update equation where $\alpha$ is the learning rate:
$
    \theta_{i} = \theta_{i} + \alpha * [r + \gamma.max_{a}Q_{\theta}(s', a') - Q_{\theta}(s, a)]f_{i}(s)
$

For the neural network based $Q$-function approximation, we trained a fully connected neural network using the same strategy as the one used to train deep Q-networks (DQN) in \cite{mnih2013playing}. The neural network given by $Q(s, a, \theta)$, where $\theta$ represents the weights of the network, approximates the $Q$-value function corresponding to the optimal policy i.e. $Q^{*}(s, a)$. The agent's transition at each time step, $\langle s, a, r, s' \rangle$ is stored in a database $D$. The database is of a fixed length and stores the most recent transitions. After each action is executed (which is treated as an iteration $i$ of the training algorithm), the Q-network is trained by optimizing the following loss function over a mini batch of transitions sampled uniformly from $D$: 
$
	L_{i}(\theta_{i}) = E_{(s,a,r,s') \sim D} [(r + \gamma max_{a'} Q(s',a'; \theta_{i-1}) - Q(s, a; \theta_{i}))^2].
$

In \name{} the Taxi Domain was trained using Model Based RL, StackBot using Approximate Q-learning and the HIV Domain using a DQN. 

\vspace{-2mm}
\section{Related Work}
\label{sec:related_work}

\subsection{Explainability in Reinforcement Learning}
\label{sec:rl_xai}
As the use of RL continues to expand, there is an increasing interest in XAI as it applies to RL. In our work, the RL agent is considered as a black box and \name{} computes explanations for its decisions. Our explanation technique is inspired from LIME~\cite{ribeiro2016should}, which attempts to explain a classifier's decisions. In contrast to LIME, which works on one-shot decision making problems, we investigate if a similar technique can explain the decisions of an RL agent which solves a sequential decision-making problem. In addition, we provide functionalities in our interface that not only answer queries about why the agent took a particular action, but also answer contrastive queries about why an agent chose a particular action over an alternative action suggested by the user, and when, in general, does an agent takes a particular action. 

There have been other works specific to explaining RL agent's policy, primarily focusing on pixel-based domains. For example, Greydanus et al.~\cite{greydanus2018visualizing} utilized saliency maps to gain insights on how an agent learns and executes a policy in a 2D video game space. Similarly, Yang et al.~\cite{yang2018learn} identified regions of interest by visualizing pixels of game images. 
Hayes et al.~\cite{hayes2017improving} proposed a set of algorithms to explain agent policies using a common modality of Natural Language (i.e text). 
Recent research~\cite{van2018contrastive} has explored policy explanation via answering contrastive queries, where an example question might look like: ``\textit{Why did the agent go right instead of going left?}'' Here, two main entities---a fact and a foil---are contrasted to explain why the fact was chosen over the foil. \name{} visualizes contrastive explanations (the ``\textit{Why not?}'' question) among the other types of questions.
(Section~\ref{sec:evaluation}).

\subsection{Visualization for XAI and RL Explainability}

For general discussion on the use of visualization for deep learning and XAI, several recent surveys are available~\cite{hohman2018visual, zhang2018visual, adadi2018peeking, choo2018visual, seifert2017visualizations}. Here, we focus on describing recent visualization tools specifically for RL analysis and explainability, which include the following:

MDPVis~\cite{mcgregor2017interactive} is a system designed for debugging and optimizing MDPs by interacting with an MDP simulator. DQNViz~\cite{wang2018dqnviz} focuses on analyzing the training of a deep RL agent, from a high-level overview down to individual epochs. DRLViz~\cite{jaunet2020drlviz} visualizes the internal memory of a deep RL agent as a way to interpret its decisions. 
Similarly, DynamicsExplorer~\cite{he2020dynamicsexplorer} is a diagonistic tool for looking into the learnt policy under different dynamic settings. 

Each of the aforementioned systems have significant differences  compared to \name{}. First, the primary objective of each tool is to support \textit{internal} debugging of a trained model, with a focus on \textit{RL expert users}. In contrast, \name{} supports \textit{RL non-experts} by promoting the human's understanding of the RL agent's exhibited \textit{external} behavior. Second, these systems are highly domain or technique-dependent, which limits their generalizability. For instance, DQNViz~\cite{wang2018dqnviz} provides a trajectory view that only supports pixel-based (specifically, Atari) video games. DRLViz~\cite{jaunet2020drlviz} only supports models trained using RNNs. In contrast, \name{} supports both model-based and model-free RL. We additionally demonstrate \name{} across three significantly different types of domains (including a safety-critical HIV domain). 

In actuality, \name{} can support \textit{any} domain provided the states can be represented in human-interpretable features. For complex domains such as Atari games where states are represented as a collection of pixels (i.e., an image), it is currently a significant open problem in the AI community to learn an interpretable representation of the pixel features that can reasonably capture the domain's dynamics. To this end, tools like DRLViz are meant for RL expert users; they sidestep the issue by showing the state directly. We provide thoughts on how to approach these types of problems in the discussion, but as this is still an open AI problem, we omit pixel-based domains from our example domains. We describe these differences to show how \name{} differentiates compared to prior tools in considering RL explainability, by focusing on different data domains and target users.

\vspace{-1mm}
\section{Pre-Study and Design Requirements}
\label{sec:dr}



To motivate a design for question-based visual explanation of agent decisions, we conducted a pre-study with three RL experts. Each had at least four years of RL research experience. This study consisted of extended email correspondences and completing a survey, all aimed at understanding the role of explanation and interaction as it relates to RL agents. Additionally, we reviewed recent papers that discuss issues of RL interpretability and transparency, which therefore provide motivation for agent explanations (e.g.,~\cite{deshpande2020interactive, amir2018highlights}). Based on the collected feedback and paper readings, we identified a set of four high-level design requirements \textbf{DR1}--\textbf{DR4}.



\textbf{DR1: Provide an overview of the state space and policy in terms of its diversity and expected future rewards.} 
RL agents might learn on domains with large state spaces whose dynamics can be modeled as networks. As these networks scale in complexity, it quickly becomes difficult for human users to understand them~\cite{yoghourdjian2018exploring}. 
Multiple pre-study participants noted that the ability to navigate and explore the state space (and the actions taken in those states) is necessary for understanding the policy of an RL agent. Further, being able to show the states and the expected future rewards (given an optimal policy) helps users identify other states that have either \textit{highly different} or \textit{highly similar} rewards. \textit{Visualization can provide an overview of the policy, with an emphasis on highlighting states with similar/different expected rewards.}

\begin{figure*}[t]
   \centering
    \includegraphics[width=0.9\linewidth]{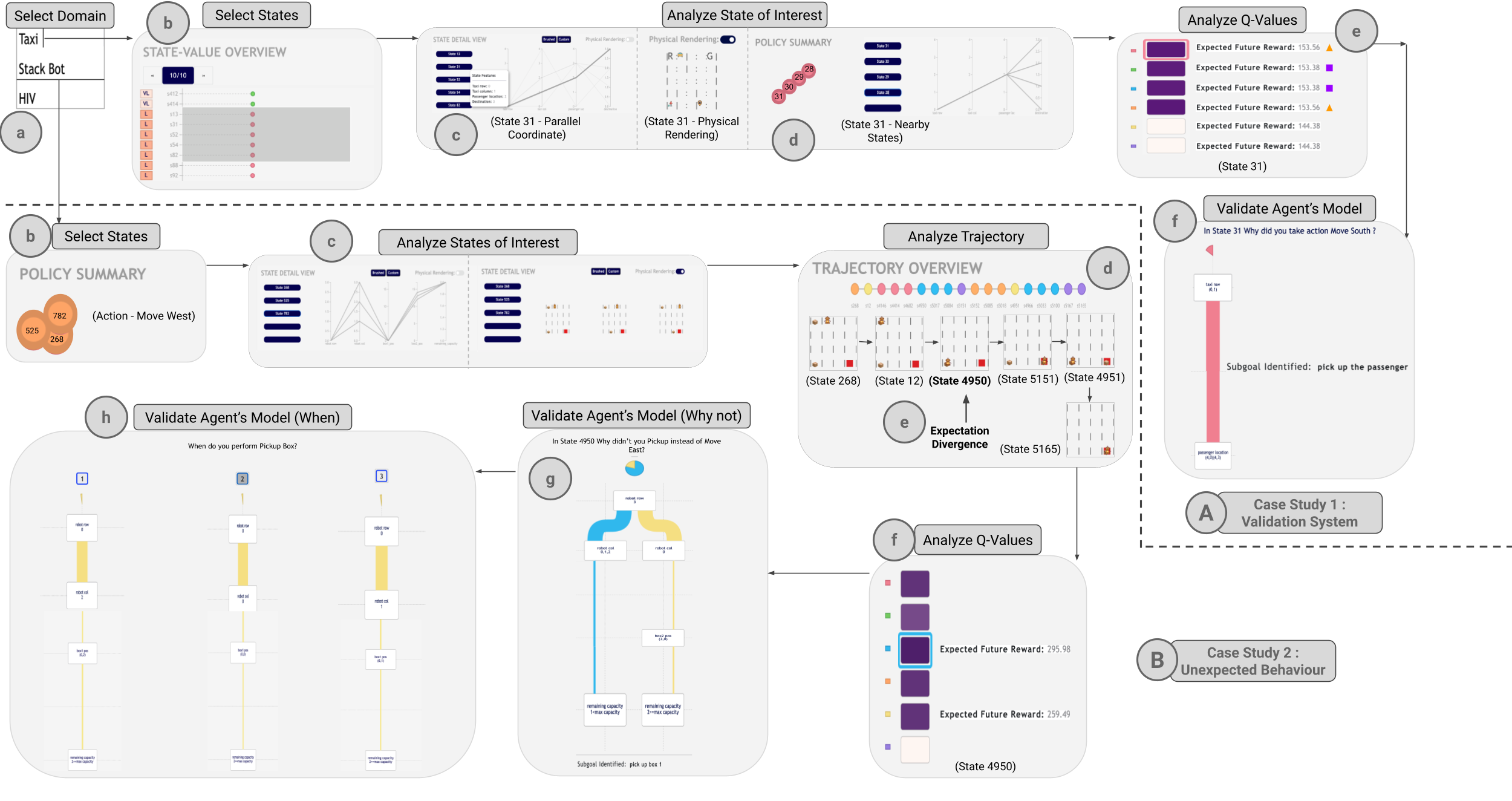}
    \vspace{-3mm}
   \caption{Gary's actions taken during the (A) case study \#1 and (B) case study \#2.}
   \label{fig:use_case}
\end{figure*}

\textbf{DR2: Provide visualizations for individual states.} Pre-study participants also discussed the importance of being able to inspect and review individual states. For humans to understand a state, it must be represented or defined in a way that provides semantic meaning. When state features correspond to a spatiophysical domain,
one solution is to represent the state using spatial visualizations, such as shown in Figure~\ref{fig:use_case}. This solution does not work, however, when states do not have a physical domain. Consider a healthcare agent for recommending a patient's treatment plan, where state features consist of a bunch of health metrics. In such cases, information visualization techniques can be employed to show the state's feature values. As opposed to providing these features in a table-based format, \textit{visualizing individual states can make it easier to analyze and compare across the potentially thousands of states that make up a policy}.

\textbf{DR3: Let the user ask questions to the agent.} While DR1 and DR2 are important to provide generalized information about the agent's policy and the individual states in the domain, they do not provide explanations or justifications for the agent's decisions. Ultimately, when a user is examining the actions taken by the agent, they will focus on questions like, ``\textit{Why} was this action taken? \textit{Why not} take this other action? \textit{When} is this action taken?'' \textit{To explain the agent's decision-making process, visualizations (and interactions) should be designed to support a question-and-answering dialogue between the humans and the RL agent.}

\textbf{DR4: Allow users to navigate the explanation space to prevent overloading.}
Finally, the explanation given by the RL agent highly depends on the state features. There might be multiple conditions in which a certain action is taken. However, giving the user all the reasons for an agent's decision (i.e., identifying every condition) might prove overwhelming. Instead, being able to identify important state regions with similar explanations, and \textit{letting user interactively choose the explanation they wish to see, can limit cognitive overhead and help in better understanding the agent's reasoning.}
\begin{figure*}[t]
  \centering
  \includegraphics[width=0.9\textwidth]{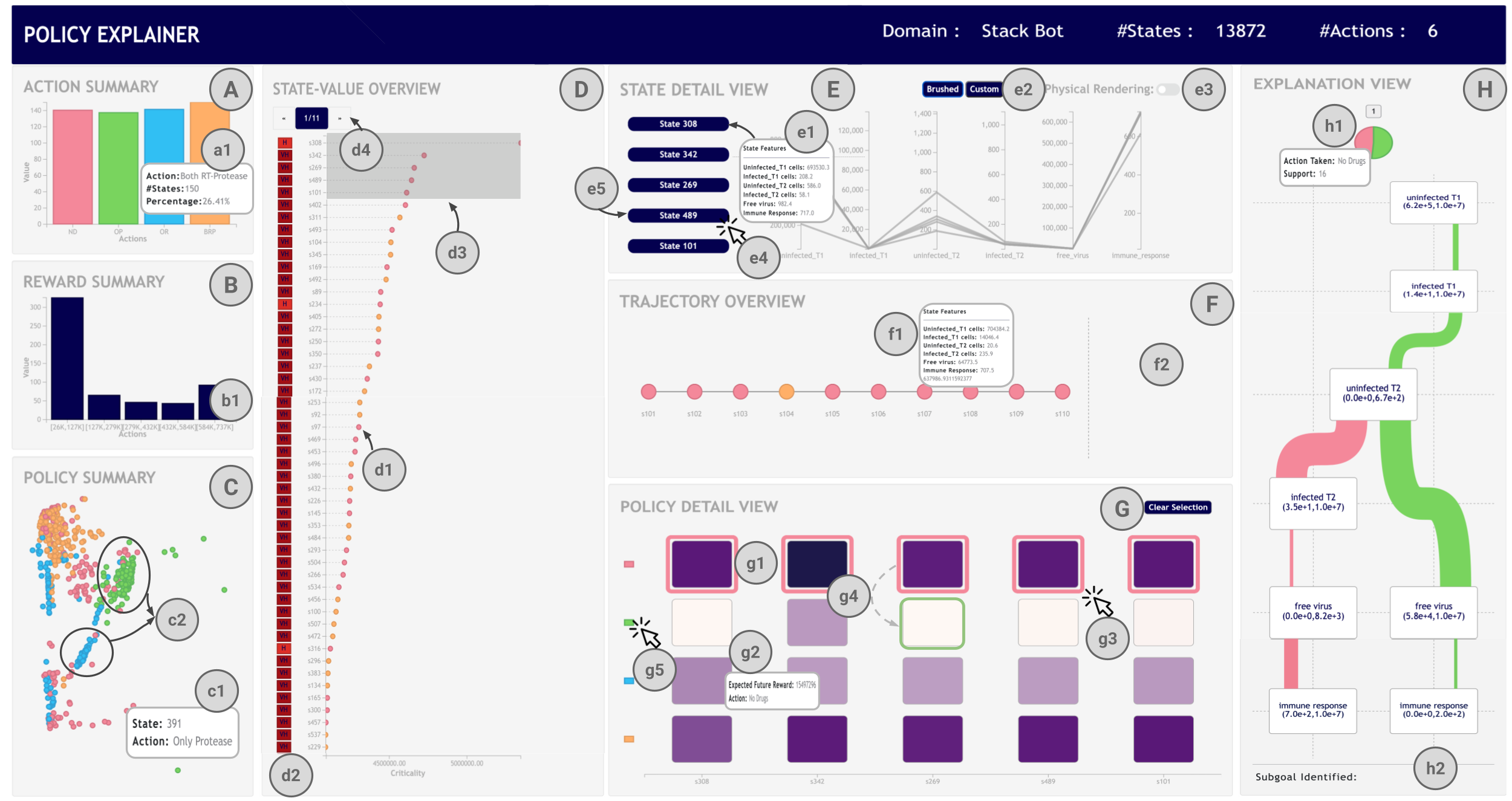}
  \vspace{-3mm}
  \caption{The \name{} interface, shown here with an HIV treatment domain~\cite{adams2004dynamic, miller2020whynot}, consists of eight main linked sections, which support (A--C)~summarizing the domain and the optimal policy, (D)~summarizing state regions and expected rewards, (E--G)~detailed analysis of states and policy, (F)~a trajectory to see the agent progression, and (H)~an explanation panel to answer \textit{Why? Why not? When?} questions.}
  \label{fig:interface}
\end{figure*}


\begin{figure}[t]
   \centering
    \includegraphics[width=0.95\columnwidth]{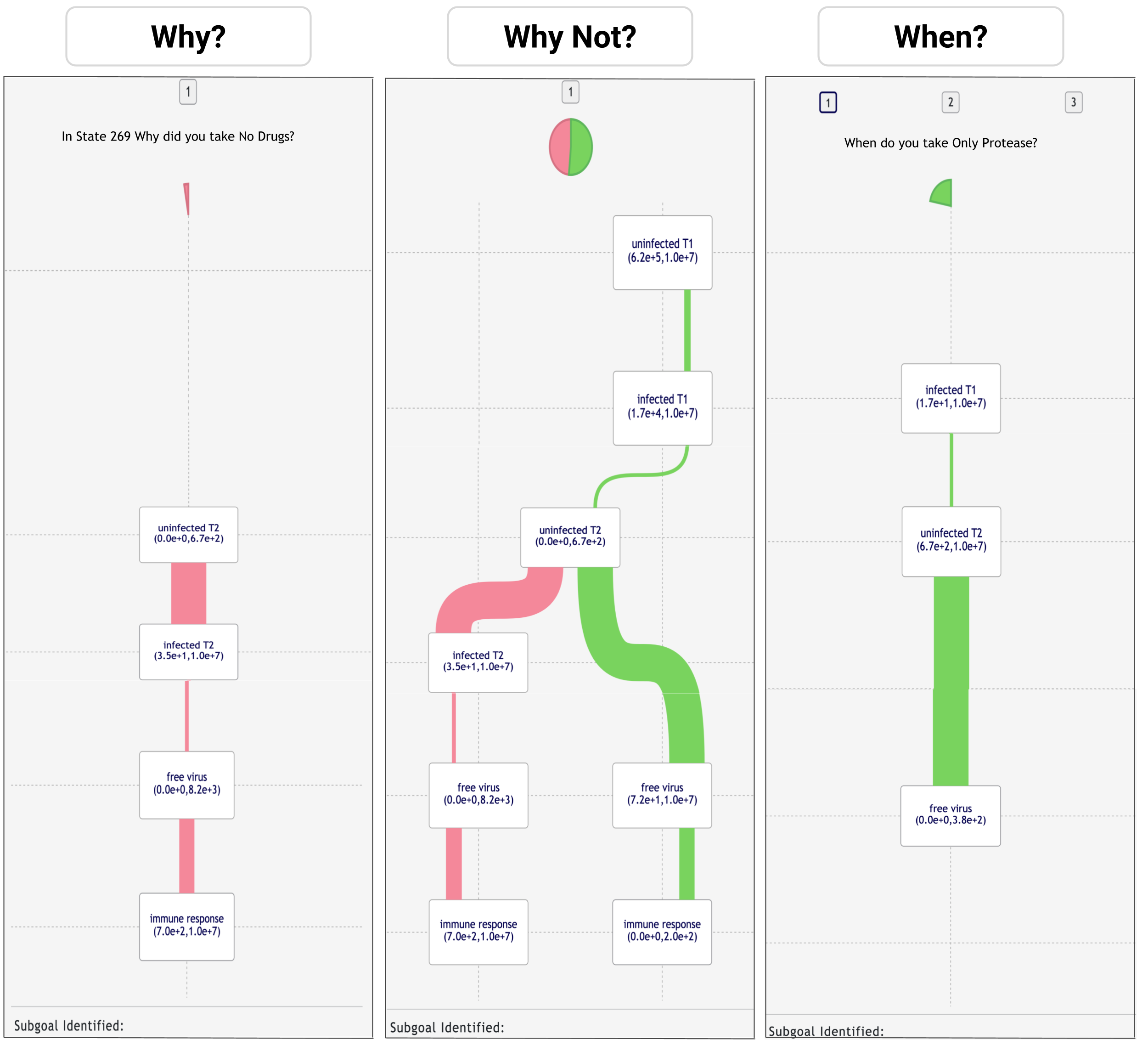}
    \vspace{-3mm}
   \caption{\name{} can answer three types of questions in its explanation panel. (\textit{Why?}) Here, the user is asking why the \textsf{No Drugs} action was taken (action colors map to the key in Figure~\ref{fig:interface}\textsf{(a1)}); there are four relevant features used in this decision. (\textit{Why not?}) Here, the contrastive question being answered is, ``\textit{Why take the (pink) \textsf{No Drugs} action instead of the (green) \textsf{Only Protease} action?}'' While one feature is shared between both actions, the rest influence the actions based on their values. (\textit{When?}) The user is asking when the \textsf{Only Protease} action is taken. The system gives top three most general reasons for this action being the optimal one. Currently, the first popular action is toggled, which shows the influence of three features.}
   \label{fig:example_explanations}
\end{figure}

\section{Policy Explanation}
\label{sec:policy_explanation}


Lim et al.~\cite{lim2009and} found that \textit{Why?} and the \textit{Why not?} are the types of questions most commonly asked to intelligent systems. Relatedly, some of the most cited papers on policy explainability (e.g.,~\cite{hayes2017improving}) highlight identifying state regions (i.e., \textit{When?} questions) as an important task. We thus focus on generating visual explanations for these three questions types. In this section we explain a methodology to generate explanations to queries submitted by a user. These explanations are used by \name{} to support interactive question-and-answer sessions between the user and RL agent.



In this work, we answer the human's question by making the agent's decision-making reasoning more transparent to the human. Informally, the idea is to highlight state features that lead to an action being chosen as the optimal action. This idea to identify the underlying features responsible to explain AI decisions has been previously explored for classification tasks~\cite{olah2017feature, dovsilovic2018explainable}. In our case, we apply this idea to explain optimal policies learned for a sequential decision making task.

To learn the salient features that effect the agent's policy, we first approximate the policy via supervised learning which learns a decision boundary based on the features that separate classes. We then extract how the classification algorithm uses these features to determine the output class. This means the algorithm must itself be interpretable. Because of this, we use a decision tree classifier to approximate the policy. A significant advantage of decision trees is that it is quite easy to track which features (and their ranges of values) lead to particular classification results. 

We now explain this approach in detail, showing how a trained decision tree classifier can be used to answer the three question types when posed from a human to an RL agent.

As explained in Section~\ref{sec:background}, each state is defined as a feature vector $\langle f_1, f_2, \dots, f_n \rangle$ and the policy $\pi^{*}$ maps each state to its corresponding optimal action. We start by using this complete mapping as a set of training samples, $\{(x_1, y_1), (x_2, y_2), \dots, (x_{|\mathcal{S}|}, y_{|\mathcal{S}|})\}$, to train a decision tree $T$, where $x_i$ represents the state features, and $y_i$ is the optimal action. The decision tree is a binary tree where each non-leaf node $n$ has some feature $f$ and a corresponding threshold value $\theta$ associated with it. The edge $e$ that connects the node to its left child represents the condition $f < \theta$ while the edge for the right child represents the condition $f \geq \theta$. We define the direction of an edge as \textit{left} / \textit{right} if it connects to the left / right child of the node. 

Any input state to the decision tree can be mapped to a unique path in $T$ from the root node to a leaf node by following the edges that the features of the state satisfy. We denote this unique path corresponding to a state $s$ as $\mathcal{P}(s)$. The leaf node would be associated with an action $a$ that should ideally be $\pi^{*}(s)$. Any path from the root to a leaf node in $T$ represents a decision rule of the form ``\textit{if} $\text{condition}_1$ \textit{and} $\text{condition}_2$ $\dots$ $\text{condition}_K$ \textit{then} $\text{action}$'' where each condition corresponds to a unique feature and it gives the range of values that feature can take for any input state to be classified as the action, while $K$ is the total number of unique features on the path. We denote the rule corresponding to a path $\mathcal{P}$ as $\text{rule}(\mathcal{P})$. Given a path $\mathcal{P}$, the associated rule can be identified by Algorithm 1. \name{} answers user queries in terms of these rules as explained in the remaining parts of this section.

An important note here is, for the domains considered in this paper, the decision tree for each case is overfitted to generate rules. This is allowed because, as we have access to the entire state space and the policy, we do not need to perform any form of testing or validation, and can focus on achieving the highest training accuracy. In each case, the fidelity of the decision tree is $>=99\%$, which means the explanations generated by the agent is highly accurate.

 \begin{algorithm}[h]
 \caption{Compute $\text{rule}(\mathcal{P})$} 
 \label{alg:alg1}
 \begin{algorithmic}[1]
 \renewcommand{\algorithmicrequire}{\textbf{Input:}}
 \renewcommand{\algorithmicensure}{\textbf{Output:}}
  \REQUIRE path $\mathcal{P}$, set of all features  in the domain $\mathcal{F}$, decision tree $T$
 \ENSURE  Rule associated with $\mathcal{P}$
 \STATE relevant\_features $\leftarrow$ \{\}
 \FOR{$f \in \mathcal{F}$}
 \STATE $f_{\text{range}}$ $\leftarrow$ set of all possible values for $f$
 \ENDFOR
 \STATE current\_node $\leftarrow$ root node of $\mathcal{P}$
 \WHILE{current\_node is not leaf}
 \STATE current\_edge $\leftarrow$ edge connected to current\_node in $\mathcal{P}$
 \STATE $C(f)$ $\leftarrow$ feature associated with the current\_node
 \STATE $C(\theta)$ $\leftarrow$ threshold associated with the current\_node
 \STATE relevant\_features $\leftarrow$ $\text{relevant\_features} \cup C(f)$
 \IF{direction of current\_edge is \textit{right}}
 \STATE current\_node\_range $\leftarrow$ [$C(\theta)$,$\infty$]
 \ELSE
 \STATE current\_node\_range $\leftarrow$[$-\infty$,$C(\theta)$]
 \ENDIF
 \STATE $C(f)_{range}$ $\leftarrow$ $C(f)_{range}  \cap \text{current\_node\_range}$
 \STATE current\_node $\leftarrow$ node adjacent to current\_node in $\mathcal{P}$
 \ENDWHILE
 \RETURN $f_{\text{range}}$ $\forall f \in $ relevant\_features
 \end{algorithmic} 
 \end{algorithm}

\textbf{Answering \textit{Why?} Questions.} These questions take the form, ``\textit{Why would you take \{action $a$\} in \{state $s$\}?}" To answer this query, we identify the specific conditions under which the action $a$ is executed by the agent. This is achieved by identifying the path $\mathcal{P}(s)$ corresponding to the state $s$ and then applying the procedure in Algorithm \ref{alg:alg1} to compute $\text{rule}(\mathcal{P}(s))$. The computed decision rule serves as the explanation. 

\textbf{Answering \textit{Why not?} Questions.} These are contrastive queries~\cite{van2018contrastive} that take the form, ``\textit{Why would you take \{action $a^{*}$\} instead of \{action $a_{f}$\} in state $s$?}'' Here, $a^{*}$ is the action chosen by the policy and $a_{f}$ is the alternate action that the user might prefer to take. To answer these type of queries, we find a state $s_{f}$ that is closest to the state $s$ where the agent would execute action $a_{f}$. The distance between any two states is calculated as the Euclidean distance between their corresponding feature vectors. We then compute two decision rules, $\text{rule}(\mathcal{P}(s))$ and $\text{rule}(\mathcal{P}(s_{f}))$, that show conditions when action $a^{*}$ and $a_{f}$ are chosen. These rules are presented to the user as a response to their query. In other words, by comparing the results of the rules, the user can assess why the policy picks $a^{*}$ over $a_{f}$. This is similar to ~\cite{hayes2017improving} where  contrastive queries are answered by comparing features of similar states under which the two actions are taken except that they use natural language for explanation. 

\textbf{Answering \textit{When?} Questions.} The final question type we support is of the form, ``\textit{When would you do take the \{action $a$\}?}'' One common motivation for this question type is for the user to understand the most frequent cases where this action is chosen. To operationalize this, we first identify all the leaf nodes in $T$ whose label is $a$. For each leaf node, we find the number of states $s \in \mathcal{S}$ whose path $\mathcal{P}(s)$ ends in that node. We pick the top three nodes with the highest number of states, and for each of these nodes, compute the path $\mathcal{P}$ from the root to that node. Finally, we compute $\text{rule}(\mathcal{P})$ for each path and use that as the response to the query.

\vspace{-2mm}
\section{Visual Analytics System : \name{}}
\label{sec:interface}

PolicyExplainer requires an agent trained using either model-based or a model-free approach. We extract data from the trained agent---not just the optimal policy, but also states, features, and the optimized Q-values and rewards for each state-action pair. If the state space is continuous, we find the most important states based on an importance function defined in~\cite{amir2018highlights}. This extracted data is fed into the \name{} interface (shown in Figure~\ref{fig:interface}), which consists of eight linked panels that provide general visual analytics about the policy and states (\textbf{DR1}--\textbf{DR2}) and lets users interactively ask \textit{Why? Why not? When?} questions to the RL agent and receive visual explanations (\textbf{DR3}--\textbf{DR4}). 

\textbf{\textsf{(A, B)} Action and Reward Summary Panel.} The action distribution and the reward distribution bar charts show summary statistics, respectively, of the frequency of actions and rewards which the agent takes or receives in its optimal policy $\pi^*$. \textsf{(a1, b1)} Hovering on the charts displays a tooltip showing exact values.

\textbf{\textsf{(C)} Policy Summary Panel.} The policy summary panel shows an overview of the optimal policy and is laid out via dimensionality reduction based on state features. Each state is encoded as a circle and the color encodes the action the agent chose. \textsf{(c1)} Hovering on the circle displays a tooltip with the state and the action chosen in the particular state. This panel along with the State value overview panel \textsf{(D)} is meant to support extracting interesting states for the user to explore. As seen in \textsf{(c2)}, the panel highlights some clusters wherein states with similar features have the same recommended action.  

\textbf{\textsf{(D)} State \& Value Overview.} 
The state value overview provides a summary of the state values over the state space (\textbf{DR1}). A horizontal lollipop chart \textsf{(d1)} showcases states arranged in a descending order of criticality. The vertical y-axis consists of individual states and the horizontal x-axis represents the critical values. A state is defined to be more critical if there is a significant difference of rewards on randomly choosing an action and the reward gained on doing an optimal action \cite{huang2018establishing}. Mathematically, this is represented with the following equation: $C(s) = max_a Q^{\pi^*}_{(s,a)} - \frac{1}{N_a} \sum_a Q^{\pi^*}_{(s,a)}$ where $C(s)$ is the criticality $C$ of state $s$, $\pi^*$ is the optimal policy, and $Q_{(s,a)}$ is the expected future reward in state $s$ on taking action $a$. The red boxes beside each state \textsf{(d2)} represent the value of the state with labels that ranges from Very High to Very Low values which are also redundantly encoded using a sequential color scale, with darker red showing higher rewards and light orange shade showcasing lower rewards. A fixed-width brush \textsf{(d3)} can be scrubbed across the state space; this selects a set of states for further analysis in the state and the policy detail panel. Similarly for exploring and navigating large state spaces we also provide users with a page navigator tool \textsf{(d4)} to do the same. 

\textbf{\textsf{(E)} States Detail View.} 
The states detail view shows a detailed visualization of states \textbf{(DR2)}. \textsf{(e5)} Selected states are shown in a blue colored panel. Each state is visualized as a line in a parallel coordinate chart, where each y-axis represents features of the state.  \textsf{(e1)} Hovering on \textsf{(e5)} shows the exact features values for that state. \textsf{(e4)} Clicking  on a state populates the trajectory panel \textsf{(F)}. \textsf{(e3)} For states that can be spatiophysically represented, a toggle can switch to this view. Apart from selecting brushed states, the user can also customize the states they wish to see by \textsf{(e2)} clicking on the customize button and entering their desired states manually.

\textbf{\textsf{(F)} Trajectory View.} This panel summarizes the ``trajectory'' of the agent starting from a selected state to an end state (either the policy goal or a user-defined goal). When a state is selected (via clicking on \textsf{(e5)}), this panel loads a simple visualization with each state represented as a circle and the action optimally chosen by the agent is encoded as the color. Each consecutive state is linked by a straight line running between them. \textsf{(f1)} Hovering on the circle shows a tooltip with feature names and corresponding values, as well as the reward gained by the agent upon taking the action. \textsf{(f2)} Space beside the visualization shows the physical animation of the trajectory the agent takes to reach the goal if the spatiophysical rendering exists.


\textbf{\textsf{(D)} Policy Detail View.}
When states are loaded in the states detail view, the policy detail is also populated. The available actions for each state are represented by a set of rectangular swatches (in the figure, each state has four available actions) colored by their Q-values if the agent takes that action. 

\textsf{(g1)} For each state, the swatch corresponding to the state's optimal action is given a colored border. The border color of this swatch corresponds to the type of action taken, using the color key from the action distribution chart in \textsf{(A)}. \textsf{(g2)} Hovering on a swatch provides details about the action and its expected reward i.e it's Q value.

To support policy explanation, this chart contains three interactions which let the user ask ``\textit{Why?, Why not? When?}'' questions to the agent \textbf{(DR3)}. \textsf{(g3)} First, clicking on the optimal action for a state asks, ``\textit{Why was this action chosen?}'' \textsf{(g4)} Similarly, dragging from the optimal action to another action for a state asks a contrastive question: ``\textit{Why not take this other action instead of the optimal action?}'' (This interaction adds a stroke on the contrastive action swatch.) \textsf{(g5)} Finally, clicking the action icons on the left side of the chart asks, ``\textit{When is this action taken?}'' Each of these questions loads the explanation panel \textsf{(H)} with the respective answer, to be described shortly.

\textbf{\textsf{(F)} Explanation Panel.} Finally, the explanation panel provides visual explanations for ``\textit{Why? Why not? When?}'' questions asked in the policy detail view \textsf{(G)}. Our approach for visual explanation is inspired from Young and Shneiderman's work visualizing boolean queries with flowcharts, where AND operations are represented as conditions on the same path, and OR conditions are shown with a forking path~\cite{young1993graphical}. For \name{}, horizontal dotted lines map to state features, which are labeled in boxes along with their corresponding possible feature values appended to the lines. A flow that links across several features (e.g., the green flow in the figure links across five features) represents an action taken based on the conjunction of several features, akin an AND operation. Links are colored by their action type using the action key in \textsf{(A)}; link thickness indicates the number of states that satisfy intermediate conditions. In contrast to Young and Shneiderman's boolean queries, policy explanation does not include the concept of OR operations. Therefore, forking indicates contrastive actions during ``\textit{Why not?}'' questions, where two actions might share some features but split on others. The use of differently colored links to differentiate the actions helps to demonstrate this. For examples of explanations for each question type, see Figure~\ref{fig:example_explanations}.

\textsf{(h1)} At the top of this panel, a pie chart shows the number of states for which the explanation holds; in other words it denotes the coverage of the explanation. A hover tooltip provides the exact counts and uses linked highlighting to show these states in \textsf{(C)}. \textsf{(h2)} For each explanation in a state, the subgoal being satisfied by the agent in the particular state is also identified if present. 

\vspace{-1.75mm}
\section{Case Studies}
\label{sec:use_case_text}
To illustrate how \name{} can be used to explore a policy and question an RL agent, we present two use case scenarios using the Taxi and the StackBot domains. 

\subsection{Use Case 1: Reassuring Users of Agent Behaviour.}

\textbf{Taxi Domain.}
The taxi domain~\cite{dietterich2000hierarchical} consists of a 5$\times$5 grid with walls separating some cells. The agent is represented as a taxi (\includegraphics[width=0.03\linewidth,height=0.03\linewidth]{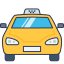})
which can move around the grid except through the walls. The grid contains four special locations: \textsf{R}, \textsf{G}, \textsf{Y} and \textsf{B}. The passenger (\includegraphics[width=0.03\linewidth,height=0.03\linewidth]{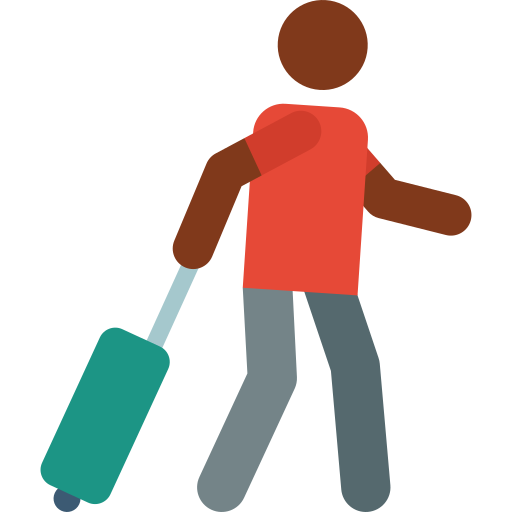}) and the destination (\includegraphics[width=0.03\linewidth,height=0.03\linewidth]{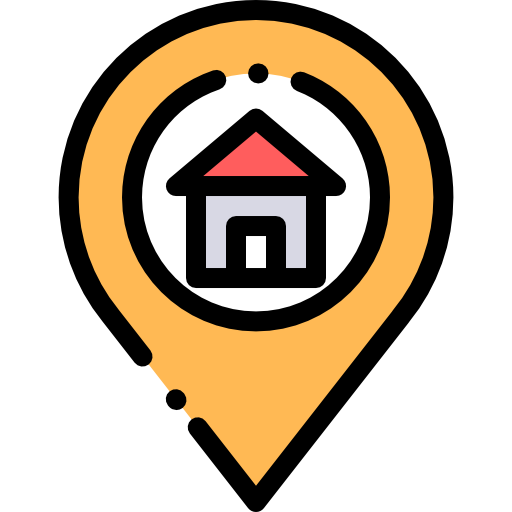}) will be present at any four of these locations. The agent's task is to pick up the passenger and drop them at the destination. Every step the agent takes entails a reward of -1; on completing the task, the agent receives a reward of +20. 

While relatively straightforward, the Taxi domain is an important and widely-used domain for demonstrating RL techniques including explainability. It is thus a good candidate domain for demonstrating \name{}'s explanation methodology and user experience.


\textbf{Gary's Analysis.}
\person{} is a human who wants to validate his understanding of the agent behaviour and make sure that the agent's understanding of the environment matches with his understanding of what actions will the agent perform. He uses \name{} to query the agent; his specific actions are shown in Figure~\ref{fig:use_case}\textsf{(A)}.

\textsf{(a)} Gary first loads the Taxi domain from the domain dropdown. This populates the action, reward summary panel and the states and values overview panel. \textsf{(b)} He uses the state-value overview to identify states of interest. He notices and selects a set of states with low expected cumulative future rewards, but with all states with the same optimal action (States 13-82). \textsf{(c)} Selecting these loads them into the state detail view. He finds State 31 to be particularly interesting, as both the passenger location and the destination are below the agent's position. \person{} wants to validate his understanding of the agent behaviour by making sure the agent first performs the \textsf{Move South} action to pick up the passenger. 

\textsf{(d)} \person{} hovers over State 31 in the state detail view and realizes that nearby states are have only a single feature (\textsf{Destination}) changing, such as States 28, 29 and 30. These states have a similar feature: the agent is north of the passenger and the optimal action in these states is \textsf{Move South}. \textsf{(e)} \person{} next looks into the policy detail view. Hovering over the tiles in the policy detail view shows an expected future reward. \person{} notices that the expected reward for moving south is higher than moving north or east, but only barely. He realizes that taking moving north or east in State 31 would require agent to return back to State 31; thus, though these actions have high rewards, they are lower than the optimal \textsf{Move South} action. This reassures \person{} that the agent will perform according to his expectations 

\person{} then clicks on the background rectangle in state 31 to ask a \textit{Why?} question: ``\textit{Why did the agent perform the Move South action in State 31?}'' \textsf{(f)} The explanation panel shows that the agent does indeed consider both its own position and the passenger's position when making decisions. The identified subgoal additionally tells \person{} that the agent is performing the \textsf{Move South} action in State 31 with the intention to pick up the passenger.

\subsection{Use Case 2: Decoding Unexpected Agent Behaviour.}

\textbf{StackBot Domain.}
The StackBot domain consists of a 4$\times$4 grid containing 2 boxes. The agent is represented as a robot (\includegraphics[width=0.03\linewidth,height=0.03\linewidth]{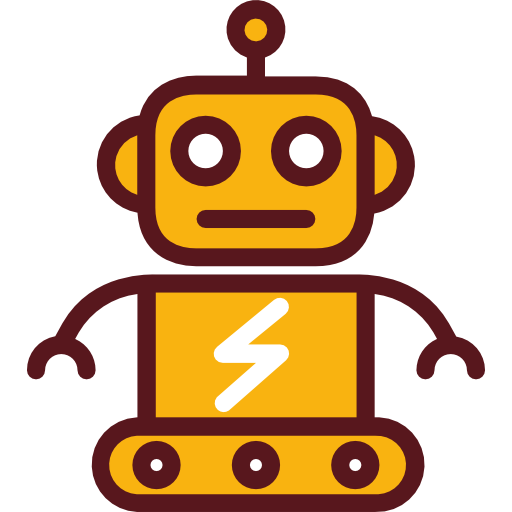}) which can freely move around the grid to pick up two boxes (\includegraphics[width=0.03\linewidth,height=0.03\linewidth]{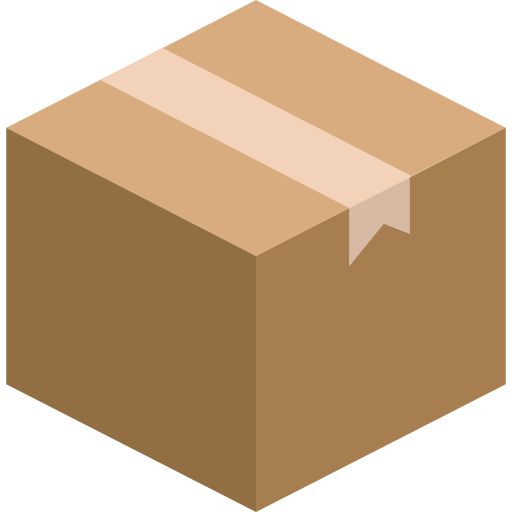}). The robot's task is to drop these boxes at the goal location~(\includegraphics[width=0.03\linewidth,height=0.03\linewidth]{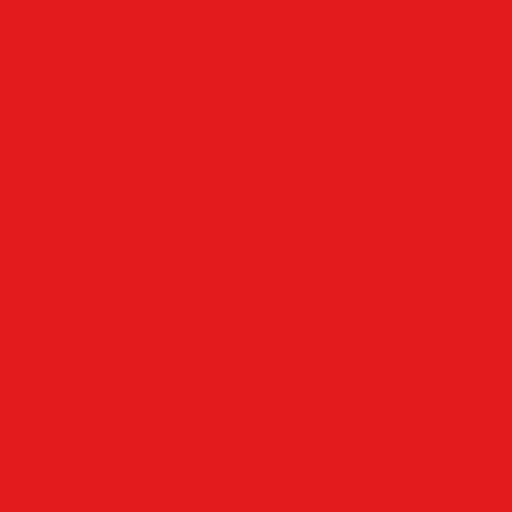}). The robot has a capacity of two boxes that it can pick up and hold at a time. However, the RL agent has trained the robot to be cautious, its policy is that it will only pick up and hold one box at a time. This counter intuitive behavior is not known to the user. Every step the robot takes gives it a reward of -1, picking up a box gives +20, and successfully dropping off a box gives +350. The episode ends when all boxes have been dropped off at the goal location, which gives  a final high reward of +500.

\textbf{Gary's Analysis.}
Gary again wants to understand the RL agent's behavior. In contrast to Case Study~\#1, the StackBot agent behaves in an unexpected manner (only holding 1 box at a time, despite having a larger capacity). Gary must resolve his gap of understanding between how he thinks the agent should act (holding 2 boxes at a time) and the agent's actual behavior. His actions are shown in Figure~\ref{fig:use_case}\textsf{(B)}; this use case is also shown in the demo video found in the supplemental materials.

Similar to the previous use case, \textsf{(a)} \person{} first loads the Stack Bot domain and reviews the action and reward summary panels, policy summary panel and the state and value overview panels. 

\textsf{(b)}
Reviewing these, he notices that some states place have the same optimal action of \textsf{Move West}. \textsf{(c)} These states have similar features: the robot column and the second box's position changing, but the robot row, first box's position and the remaining capacity remains the same. \textsf{(d)} \person{} reviews the trajectory of State 268 (i.e., the actions taken from here to finish the task) and realizes that the agent could have picked up a second box but did not (i.e., it was at that box's location and had capacity). Instead, it continued moving to the goal location to drop off the single box it was carrying. \textsf{(e)} \person{} identifies the state where the unexpected action of \textsf{Going East} happens instead of the expected action \textsf{Pickup Box} as State 4950. \textsf{(f)} He loads that state into the policy detail view to analyze its Q-values. The Q-value boxes reveal a minor dip in the expected future reward if the agent chose to pickup the box, which is counterintuitive to \person{}'s mental model of how the agent is supposed to work.

To gain more insight into this unexpected behaviour, \person{} asks a contrasive question to the agent: ``\textit{Why did the agent perform action Move East instead of the action Pickup Box in State 4950?}'' \textsf{(g)} The explanation informs \person{} that the agent moves east whenever any one of the box positions is to the left of the goal position (located at box (3,3)) and the remaining capacity is 1 (i.e., the robot is holding a box). 
On seeing this contrastive explanation, \person{} notices that the robot picks up boxes when it is at the same location and its capacity is 2 (i.e., it is not holding any boxes). \person{} wants to see if this behavior is always true regardless of the location, so he asks a \textit{When?} question to the agent: ``\textit{When does the agent perform action Pickup Box?}'' \textsf{(h)} The explanation shows that a pickup happens only when the robot is not holding a box (specifically, when its remaining capacity is equal to its maximum capacity). \person{} now understands why the agent dropped of the first box rather than picking up the second box. 

\vspace{-1.75mm}
\section{Evaluation}
\label{sec:evaluation}

To empirically evaluate \name{}, we conducted two studies: a controlled usability study with ten graduate computer science students (p1--p10), and extensive usability reviews with three domain experts who research HIV and vaccinology (e1--e3). Notably, \textit{none of our participants were experts in RL}. These evaluations serve two purposes: (1)~To understand how \name{}'s visual representations and question-and-answer dialogues support understanding of a learnt policy by non-experts. (2)~To compare \name{}'s visual explanation approach against a state-of-the-art text-based explanation baseline. 


\begin{figure}[t]
   \centering
    \includegraphics[width=0.87\columnwidth]{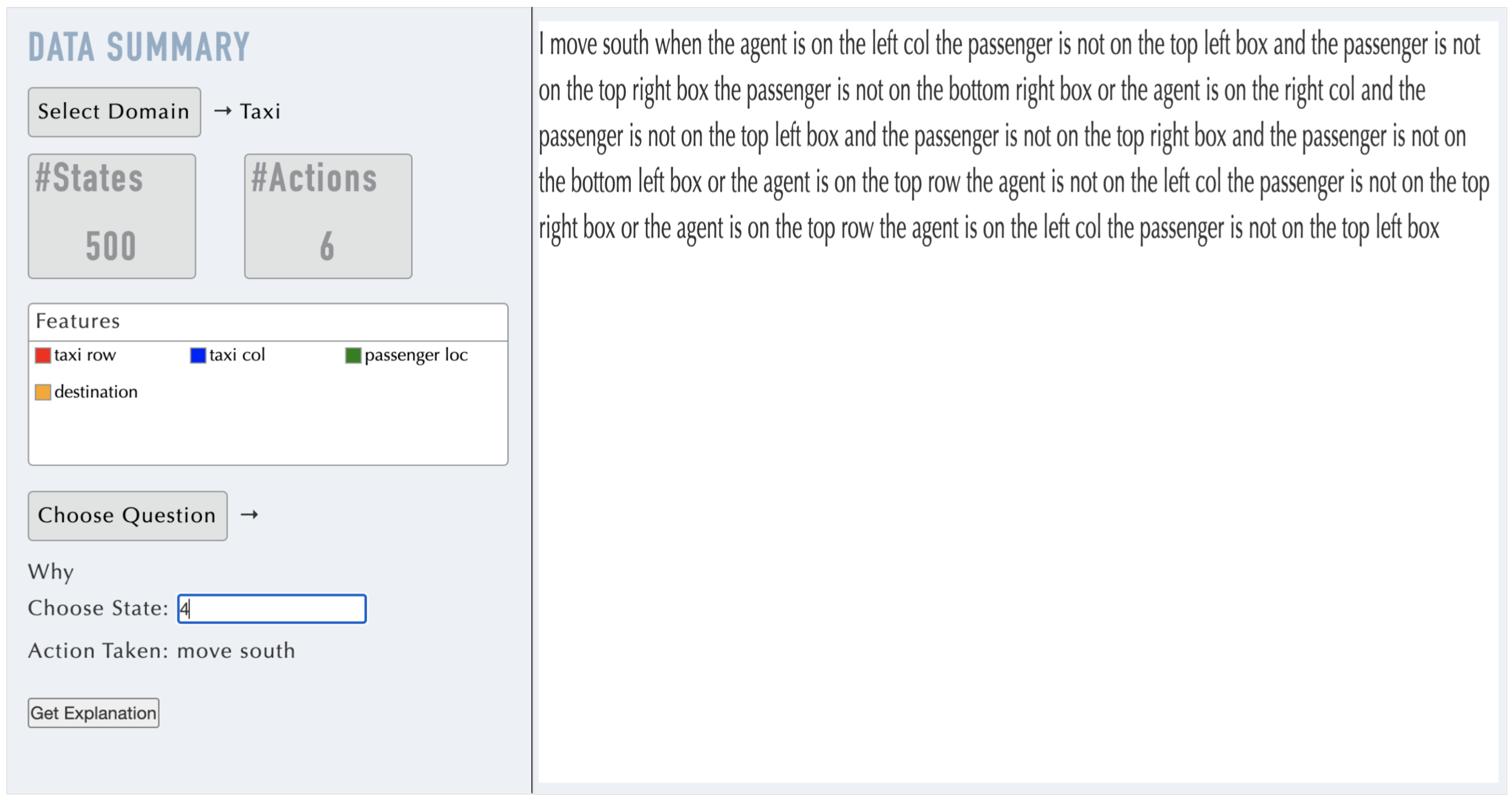}
    \vspace{-3mm}
   \caption{The baseline interface created for the user studies, which employs text-based explanations from~\cite{hayes2017improving}.
   }
   \label{fig:baseline_interface}
\end{figure}

\subsection{Study \#1 Design and Setup}

\textbf{Baseline.} As a baseline for comparing against \name{}'s visual explanations, we utilize the text-based policy explanation technique from~\cite{hayes2017improving}. Despite being a relatively recent publication, this is one of the most cited papers for policy explanation via natural language generation, and is still considered a state-of-the-art approach in the community. This technique supports ``\textit{Why?}'' and ``\textit{Why not?}'' questions as well as understanding situational behaviour, making it analogous to our three question types supported by \name{}. Explanations are based on the software's control logic and a user query; the output consists of all possible conditions for the asked question. Unfortunately, like most text-based explanation approaches, our assumption is that even for simple domains (like Taxi) this approach can quickly lead to long run-on sentences that are cumbersome for humans to parse through (see \textbf{DR4}). One motivation for \name{} is that visualization can potentially both improve interpretation and alleviate cognitive load by representing explanations in easy-to-understand and interactive visual encodings.

We downloaded the GitLab code for this technique~\cite{gitlab} and tweaked the code to obtain answers for the ``\textit{Why? Why not? When?}'' question types discussed in this paper. We then created a simple frontend interface to support the user interactively querying an RL agent, shown in Figure~\ref{fig:baseline_interface}.

\textbf{Domains.}
Three domains were used in Study~\#1: the Taxi and StackBot domains described in Section~\ref{sec:use_case_text}, and a safety-critical HIV drug treatment domain. The HIV domain~\cite{adams2004dynamic} consists of an RL agent recommending a drug cocktail for a patient. States in this domain consist of six features: (\textsf{uninfected CD4+ T-lymphocytes}, \textsf{infected CD4+ T-lymphocytes}, \textsf{uninfected macrophagus}, \textsf{infected macrophagus}, \textsf{free virus}, \textsf{immune response}). Four actions are available to the agent: (\textsf{No drugs}, \textsf{Only Protease}, \textsf{Only RT}, \textsf{Both Protease and RT}). Based on the consequences of an action in a given state (either positive or negative), the agent is given a positive or a negative reward.

\textbf{Design.} 
The study design consisted of five stages: 

\textit{(1) Interface Assignment and Training Stage.} 
First, the participant was assigned one of the interfaces. A hands on training was given, explaining available system features and interactions. Participants could ask questions and play around with the interface until they felt comfortable enough to proceed. 

\textit{(2) Task Stage.} 
For this stage, participants were shown explanations provided by the assigned interface for the three types of supported questions: \textit{Why?}, \textit{Why not?}, and \textit{When?}. Participants were tasked to rate the explanations for the question type, which we refer to as tasks t1, t2, and t3, respectively. In \name{} the users could use visualizations from other panels to aid their understanding of the explanation generated. 

Participants completed three trials for each task, or nine total trials. Tasks were timed; participants were told to notify the administrator if or when they thought they understood the agent's explanation for its decision. For each task, we had pre-selected six states from the Taxi domain; three were chosen for the participant's assigned interface and the others were held out. At the completion of the nine trials, participants completed a short survey rating the understandability of the explanations based on a 7-point Likert scale.

\textit{(3, 4) Training and Task Stages with the other Interface.} After completing the task stage with an initially-assigned interface, participants repeated the training and task stages with the other interface. Trials in the second iteration of the task utilized the states that were held out in the first iteration. To minimize potential confounds, the order of interface assignments, the selection of states for each interface, and the trial ordering was counterbalanced among participants.



\textit{(5) Freeform Analysis Stage.} 
Finally, participants opened \name{} and could freely explore the three domains. No specific task was assigned in this stage, but participants were encouraged to put themselves into the following scenario: They are a supervisor in a company with autonomous agents employed and were told to report back any unexpected situations that occur, with their understanding of the agent's reasoning. In this stage, we wanted to understand how \name{}'s features and overall user experience support RL interpretability, so the baseline interface was not used. Participants had ten minutes to complete this stage, and utilized think aloud protocol to verbalize their cognitive processes. At the end of the stage, participants completed a short usability survey and, if desired, could provide additional commentary about \name{} and baseline.

\textbf{Participants and Apparatus.} Ten graduate computer science students were recruited from $<$\textsf{Anonymous University}$>$ (average age $= 24.6$, SD $= 1.42$; 7 males, 3 females). Although some of the graduate students were familiar with AI/ML, all reported little-to-no experience in RL. Each session lasted between 45--60 minutes. 

During study sessions, both interfaces were shown in Google Chrome in full screen mode at $3840\times2160$ resolution. Sessions were held in a quiet, office-like environment with no distractions.

\subsection{Study \#1 Results}

Where applicable, we report Mann-Whitney U tests to indicate if there is a statistical difference in explanation understandability between \name{} and baseline (using a threshold of $p = 0.05$) by providing \textit{U} and \textit{p} values. 

\subsubsection{Task Stage Performance}

Two types of data points were collected during task stage trials: the time taken to complete each trial, and ratings from the participant about the understandability of explanations from each interface.

For \name{}, the average completion time in seconds for each task was t1$= 39.7$, t2$=58.5$, and t3$= 56.2$. While we planned to compare the task times between the two interfaces, there were several instances of baseline trials where participants gave up trying to understand the explanation, for all three question types, with the justification that the explanations were too verbose and difficult to understand. Participant \textit{e10} succinctly stated this: ``\textit{The text explanations were hard to understand, I felt like giving up while doing the tasks since it was too much of mental effort.}'' Thus, while the completed text explanation trials took more time on average (t1$= 97.8$, t2$= 86.9$, t3$= 55.8$), this data is skewed because many baseline trials were simply given up.

\begin{figure}[t]
   \centering
    \includegraphics[width=0.85\columnwidth]{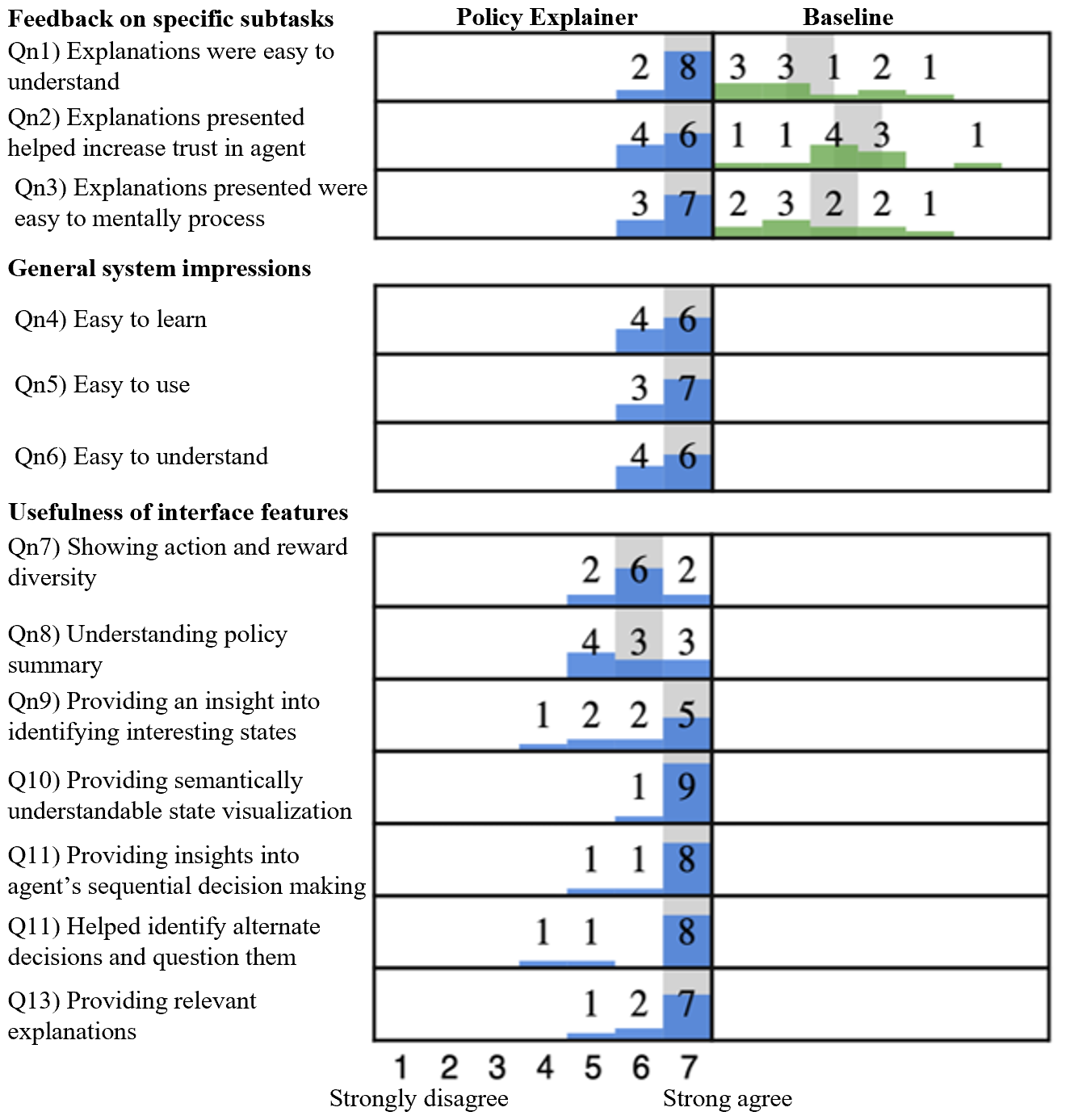}
    \vspace{-3mm}
   \caption{Participants’ ratings about various system aspects after the Freeform Stage. Median ratings are indicated in gray.}
   \label{fig:study1_ratings}
\end{figure}

Three questions were asked to participant about the understandability of the explanations for the agent behavior, shown in Figure~\ref{fig:study1_ratings}(Qn1--Qn3). For each question, \name{} performed significantly better in terms of understanding $(U = 0 , p < 0.005)$, gaining trust $(U = 2, p < 0.005)$ and the cognitive effort required to understand presented explanations $(U = 0, p < 0.005)$. These results indicate that participants felt \name{}'s visual explanations were much easier to understand compared to the baseline's state-of-the-art text explanations.

\subsubsection{Freeform Stage: User Comments and Survey Ratings}

Here, we report comments and feedback collected during and after the freeform analysis stage. Figure~\ref{fig:study1_ratings}(Qn4--Qn13) shows participant survey feedback about using the system during this stage. \name{}'s functionality and interface features were highly rated by almost all participants. Since the baseline was not used in the freeform stage, it does not have corresponding ratings for these questions, though several participants compared the two interfaces during and after this stage.

\textbf{Visual explanations were preferred to text-based explanations.} All ten participants preferred the visual explanations provided by \name{} over the baseline's text explanations. One possible reason for this, referenced by four subjects (p3, p7, p9, p10), is that \name{}'s visual explanations were more succinct compared to the verbose text explanations from the baseline. ``\textit{The text explanations were hard to understand. However, the visual one was better since the \ldots explanations were succinct}'' (p10). The idea that the text explanations explicitly required more mental effort was a common theme, stated by three participants (p4, p9, p10). ``\textit{The text explanations were hard and made no sense so I gave up}'' (p4). ``\textit{This work on visualizations for explainability makes more sense than text-based systems because the mental effort in the latter is too much.}'' (p9). These comments echo the Likert score ratings in Figure~\ref{fig:study1_ratings}(Qn1--Qn3).

\textbf{Decoding agent behaviour across panels.} 
\name{} contains several panels (apart from the explanation panels), which several participants mentioned they used to contextualize the agent's reasoning better and validate their understanding of the agent behaviour. For examples, four participants (p2, p4, p7, p10) found the Q-values for certain actions intuitive, despite being RL non-experts. ``\textit{Like in state 4, its really interesting to see the rewards for move south and move west are the same, since the agent would complete the task in the same number of steps, but if you look at the Q value for move west and north though they are same its a bit lower than the optimal action because of taking an extra step but still being in the same location}'' (p7). ``\textit{Q values had encoded rewards which aligned with our expectation of the agent }'' (p2). 

Two participants (p9, p10) especially liked that the state value overview arranged states by criticality. ``\textit{In state 276, the taxi already had a passenger so doing anything else that takes it away from the goal position, thus will have a more negative reward than the optimal action}'' (p9).
Another participant commented how the features presented in the agent explanation mimic what a human would think if they were the agent. ``\textit{Here in this explanation of why the agent moved south, it's interesting to see that the agent doesn't look at the destination feature before it picks up, it's just its own position and passenger position. Once it does, it then looks at the destination location. This thinking sort of aligns in the way humans would think which is interesting to see in a robot}'' (p4). 

All ten participants found the state and trajectory visualizations, along with the subgoal, to be particularly helpful in understanding and validating the explanations. ``\textit{The subgoal is easy but interesting to see, since it tells us what the robot is trying to achieve and I can easily validate it from the trajectory view. Same goes with the state visualizations, they are easy to understand and easily help me to understand the explanation which is way harder in the text based format}'' (p4).

\textbf{Unexpected behaviour in StackBot domain made sense on asking questions.} All ten participants used the StackBot domain, and all ten considered it to be the most interesting domain, since the agent performed counter intuitive actions. The system's explanations helped them to understand the agent's reasoning. 
``\textit{The questions I asked in StackBot helped me see the application of this interface better. It shows a very clear use of the system. Especially the `Why not?' question}'' (p1). 

\textbf{On-demand training to improve usability.} Though overall of the users found \name{} easy to learn, use, and understand (Figure~\ref{fig:study1_ratings}(Qn4--Qn6)), two participants (p3, p4) mentioned that additional training time could help them more intuitively understand the system's functions and improve the user experience. Each suggest including on-demand user guides and tutorials. ``\textit{Once you explained the interface I found the visualizations easy. Some sort of tutorial is needed though to understand the interface. }'' (p3). ``\textit{Maybe you could add a tutorial for users for the interface.}'' (p4). Fortunately, this functionality was not necessary during the study, as the administrator was present to assist participants if they were stuck or confused.

\subsection{Study \#2: HIV Domain Experts}

For Study~\#2, we worked with three researchers (e1-e3) over several weeks to evaluate and assess \name{}. These domain experts research HIV and vaccinology, particularly in low-income countries. Each has at least 4 years research experience, but none had technical familiarity about RL. Communication included several emails and videoconference interviews, as well as pair analytic sessions with both \name{} and the baseline interface, primarily using the safety-critical HIV domain (though the other domains were also demoed).

\textbf{Pair Analytics.} Notably, \textit{pair analytics}~\cite{elmqvist2015patterns, arias2011pair} is an established method for visualization evaluation by capturing reasoning processes in visual analytics. To evaluate an interface such as \name{}, a visualization expert well-versed with the system functionality ``drives,'' while the study participant freely makes analysis and investigative decisions based on their own expertise and desires. Freeform verbal discussion between the driver and participant is the basis for understanding of the participant's sensemaking process as well as what specific insights are uncovered during investigation.


\textbf{Domain Expert Feedback}
The domain experts had several comments pertaining to the explanations about HIV treatment, primarily based on their previous clinical experiences. 
All the three experts found \name{} easy to use and understand, particularly compared to the baseline. ``\textit{The interface is easier to understand}'' (e1). Likewise another professor noted that, ``\textit{It's streamlined and clean. Very spacious and clear to look at actually}'' (e2). 
One expert particularly liked the policy cluster panel: ``\textit{There are some real nice clusters here and State 1 is an outlier. Very interesting. This cluster does suggest that the agent did learn giving same drugs to similar patients} ''(e2).  

Two of the three experts noted that the question-and-answer dialogues were helpful in letting them think through and analyze diverging view points. As one commented, ``\textit{This interface also helps ask questions so it actually makes us think and check about the other possibilities. Especially the Why Not [question]} ''(e3).


In terms of expanding \name{} to work with safety-critical domains, one participant suggested incorporating user feedback into the system which updates the policy, thus making the agent adapt to the patient's needs. ``\textit{Can you change the agent behavior and the agent adapts based on the physician’s response? Not sure how hard it is but an adaptive interface would be so interesting. This aligns with something we call as intervention consideration for different patients}'' (e2). 
Another expert suggested showing confidence values instead of Q-values. 
Another expert also suggested such a system could be effectively tailored for clinicians in rural and poor settings: ``\textit{AI agents and visualizations are not that abundant in healthcare. This interface can be used as an assistive agent in low income countries}'' (e3). In such settings, where users would likely have less technical proficiency and visualization literacy, additional narrative cues could be employed: 
``\textit{Adding more annotations to the visualizations and the interface would be easier for us who don't know computer science}'' (e2).

\vspace{-1.75mm}
\section{Discussion, Limitations and Future Work}
\label{sec:discussion}
We view \name{} as a first attempt to make a generalizable visual interface to support interactive querying and explanation of an RL agent. Here, we discuss how the process of developing and evaluating \name{} demonstrates how visualization-based approaches can be leveraged to decode the behavior of autonomous RL agents, particularly for RL non-expert users. We also discuss some of the current assumptions and limitations the system makes, and how they can be improved in future research efforts.

\textbf{\name{} effectively showcases real-world and safety-critical domains.} Each of the three tested domains represent reasonable RL problems; the Taxi domain, while simple, is used for demonstrating RL techniques, and the HIV and StackBot domains mimic ``real world'' problems in terms of complexity. In particular, the HIV domain represents a safety-critical domain where model recommendations require justification and interpretability to be trusted. The domain experts in Study~\#2 found \name{}'s explanations made sense and were considered informative, indicating that such systems can effectively promote trust and interpretability for users who are not experts in RL.



\textbf{Scaling to large state spaces.} One current limitation for \name{}'s interface is that it has limited state and action scalability. For example, the HIV domain is technically a continuous state space; we convert it to a discrete state space by extracting the 568 most important states encountered by the agent. Some RL applications, such as automating video games (Atari, Pac-Man, etc.), scale up to millions of state spaces. One way to improve generalizable visual analytics explainers for RL is to accommodate RL applications with large (or truly continuous) state spaces. One strategy for this, which we intend to pursue as a future work, is to devise approximation and importance functions which extract or cluster important states. Dimensionality reduction techniques are another strategy, by reducing to a smaller number of embedded states which can then be analyzed.


\textbf{Interpreting features.} 
In Section~\ref{sec:interpresting_features}, we describe how we assume the state features that define a MDP are easily understandable by a human user. As a future work, we plan to extend this system by relaxing that assumption. One strategy is providing access to a mapper of states to features, that a human-in-the-loop can understand and update. We plan to explore the technique presented in~\cite{NEURIPS2019_77d2afcb}, which automatically learns the important human interpretable concepts that can be used to explain a neural network based classifier decisions. 

\section{Conclusion}
\label{sec:conclusion}

\name{} is a visual analytic system that lets a human query an RL agent, which answers by creating explanatory visualizations. \name{} supports both model-based and model-free RL across a variety of domains, and represents a first step in questioning-and-answering an RL agent via interactive visualization, particularly for RL novice users. This paper provides a policy explanation methodology that can be used to create succinct visual encodings for \textit{Why? Why not? When?} questions. Results from a controlled user study found these visual encodings preferable to text explanations created by natural language generation. Additionally, visual explanations via \name{} were found to increase trust in the decisions given by an RL agent, including in domain experts for a safety-critical healthcare domain. Future work intends to expand on the types of RL domains that can be handled by visual analytics systems, including domains with extremely large state spaces and features that are not human-interpretable.

\bibliographystyle{abbrv-doi}

\bibliography{template.bib}
\end{document}